\documentclass[runningheads]{llncs}
\usepackage[T1]{fontenc}
\usepackage{tabularx}
\usepackage{array}
\newcolumntype{x}[1]{>{\centering\arraybackslash\hspace{0pt}}p{#1}}
\usepackage{enumitem}
\usepackage{booktabs}
\usepackage{tikz}
\usepackage{adjustbox}
\usepackage{amsmath}
\usepackage{listings}
\usepackage{hyperref}
\usepackage[capitalise]{cleveref}
\usepackage{comment}
\crefname{claim}{Claim}{Claims}
\usepackage[position=b]{subcaption}
\usepackage[many]{tcolorbox}
\usepackage{enumitem}
\setlist[enumerate]{itemsep=0pt}

\newif\ifcomments
\newif\ifchanges
\commentsfalse\changesfalse
\commentstrue\changestrue %

\RequirePackage{adjustbox}

\RequirePackage{xspace}
\RequirePackage{soul}
\RequirePackage{xcolor}
\newcommand{\df}{\ensuremath{\mathrel{\smash{\stackrel{\scriptscriptstyle{
    \text{def}}}{=}}}} \;}
    
\newcommand  {\myclass} [1]  {\ensuremath{\textsf{\upshape #1}}}

\newcommand{\StaClass}[1]{\myclass{#1}\xspace}

\newcommand{\CQ}[1][]{\StaClass{CQ}}
\newcommand{\UCQ}[1][]{\StaClass{UCQ}}
\newcommand{\CQneg}[1][]{\StaClass{CQ\ensuremath{^{\mneg}}}}
\newcommand{\UCQneg}[1][]{\StaClass{UCQ\ensuremath{^{\mneg}}}}

\providecommand {\calC}      {{\mathcal C}\xspace}
\providecommand {\calD}      {{\mathcal D}\xspace}

\providecommand {\calN}      {{\mathcal N}\xspace}

\providecommand {\calP}      {{\mathcal P}\xspace}

\providecommand {\calS}      {{\mathcal S}\xspace}

\providecommand {\calV}      {{\mathcal V}\xspace}

\ifcomments
\newcommand{\commentbox}[1]{\noindent\framebox{\parbox{0.98\linewidth}{#1}}}

\newcommand{\acomment}[2]{\ \\ \fbox{\parbox{0.98\linewidth}{{\sc #1}: #2}}}
\newcommand{\mcomment}[2]{{\color{blue}(#1)}\footnote{#1: #2}} %
\else
\newcommand{\commentbox}[1]{}
\newcommand{\mcomment}[2]{}
\newcommand{\acomment}[2]{}
\fi

\ifchanges

\setul{}{0.2mm}
\setstcolor{red}

\else

\fi

\usepackage{graphicx}
\usepackage{amssymb}
\usepackage{color}

\newcommand{\nonterminal}[1]{\ensuremath{\langle\mathtt{#1}\rangle}}
\newcommand{\termi}[1]{\text{``#1''}}

\usetikzlibrary{arrows}
\pgfdeclarelayer{backgroundA}
\pgfdeclarelayer{backgroundB}
\pgfdeclarelayer{background}
\pgfdeclarelayer{foreground}
\pgfsetlayers{background,backgroundB,backgroundA,main,foreground}

\usetikzlibrary{shapes.misc}
\usetikzlibrary{decorations.text,decorations.footprints,decorations.shapes,shapes.symbols}
\pgfdeclarelayer{background}
\pgfdeclarelayer{foreground}
\pgfsetlayers{background,main,foreground}
\usetikzlibrary{matrix,calc,positioning,backgrounds,fit}
\usetikzlibrary{shapes.multipart,shadows,shadows.blur}

\newcommand{\Mstrut}{\vrule height 0.4cm depth 0.1cm width 0pt}

\newcommand{\taskcontent}[3]{\vspace{-1ex}\ifthenelse{\equal{#1}{}}{}{\textbf{#1: }}\textbf{#2}\ifthenelse{\equal{#3}{}}{}{\\[-0.5ex] \Mstrut #3}\vspace{-0.5ex}}
\newcommand{\smalltaskcontent}[5]{%
    \vspace{-1ex}\ifthenelse{\equal{#1}{}}{}{\textbf{#1: }}\textbf{#2}%
    \ifthenelse{\equal{#3}{}}{}{\\[-0.5ex]\textbf{\Mstrut #3}}%
    \ifthenelse{\equal{#4}{}}{}{\textbf{}\\[-0.5ex] \Mstrut #4}%
    \ifthenelse{\equal{#5}{}}{}{\\[-0.5ex]\Mstrut #5}%
    \vspace{-0.5ex}}

\definecolor{iltisheader}{rgb}{1,.89,.635}
\definecolor{iltistask}{rgb}{1,.95,.816}
\definecolor{iltisedge}{rgb}{.945,.725,.333}
\definecolor{iltispopupheader}{rgb}{.75,.675,.607}

\definecolor{iltispurpel3}{rgb}{.522,.471,.651}
\definecolor{iltispurpel4}{rgb}{.435,.373,.588}
\definecolor{iltisblue0}{rgb}{.89,.95,1}
\definecolor{iltisblue1}{rgb}{.8,.91,1}
\definecolor{iltisblue2}{rgb}{.55,.72,.85}
\definecolor{iltisgrey1}{rgb}{.93,.92,.91}
\definecolor{iltisgrey2}{rgb}{.85,.84,.82}
\definecolor{iltisgrey3}{rgb}{.76,.75,.74}

\newcommand{\borderradius}{1pt}
\newcommand{\borderwidth}{.8pt}

\tikzset{
	task/.style={
		rectangle split,
		rectangle split parts=3,
		rectangle split part fill={iltisblue1, iltisblue0, iltisblue1},
		rectangle split part align={left, center,left},
		rectangle split draw splits=false,
        every one node part/.style={font=\scriptsize},
        every two node part/.style={font=\footnotesize},
        every three node part/.style={font=\scriptsize},
		rounded corners=3pt,
		inner xsep=5pt,
		inner ysep=4pt,
		align=center,
		line width=.3pt,
    },
	taskedge/.style={
        ->,
        draw,
        line width = 2pt,
        iltisblue2,
        shorten <=2pt,
        shorten >=2pt
    },
	exercise/.style={
        node distance = 0.3cm,
		every node/.style={task},
		every edge/.append style={taskedge},
	},
    connector/.style={
        decoration={footprints,foot of=felis silvestris,foot length=5pt,stride length=10pt,foot sep=1pt},
        decorate,
        iltisedge,
    },
    comment/.style={
		draw=iltisheader,
		fill=iltistask,
		rectangle,
        rounded corners=\borderradius,
		line width=\borderwidth,
		inner xsep=5pt,
		inner ysep=4pt,
		node font=\scriptsize,
	}
}

\newcommand{\includeScreenshot}[6]{
	\includegraphics
	[%
	width=#2,%
	clip,%
	trim=#3 #4 #5 #6,%
	]%
	{#1}%
}

\tikzset{screenshot/.style={
    inner sep=0,outer sep=0,
    align=center, %
}}

\tikzset{screenshot border/.style={
    draw=iltisheader,
    rounded corners=\borderradius,
    line width=\borderwidth,
    inner sep=0,outer sep=0,
    inner xsep=-1.5pt,
    align=center, %
}}

\newcommand{\sizeof}[1]{\ensuremath{|#1|}}
\newcommand{\noComparisons}{\ensuremath{\sizeof{\calV}}}
\newcommand{\condPos}{\ensuremath{P[\calP\!\mid\!\text{pos.}]}}
\newcommand{\condNeg}{\ensuremath{P[\calN\!\mid\!\text{neg.}]}}
\newcommand{\accumulated}{A}
\newcommand{\methodB}{(1) \texttt{B}}
\newcommand{\methodBGP}{(2) \texttt{B+G+P}}
\newcommand{\methodBGO}{(3) \texttt{B+G+S}}
\newcommand{\methodBSP}{(4) \texttt{B+S+P}}
\newcommand{\methodBSO}{(5) \texttt{B+S+S}}
\newcommand{\methodGPT}{(6) \texttt{gpt}}

\newcommand{\anonymize}[1]{#1}
\newcommand*{\Iltis}{\anonymize{Iltis}\xspace}

\begin{document}
\title{Logical Modelling in CS Education: \\ Bridging the Natural Language Gap}

\titlerunning{Logical Modelling in CS Education: Bridging the Natural Language Gap}
\author{
\anonymize{Tristan Kneisel\inst{1}\orcidID{0009-0001-9160-4330}}
\and
\anonymize{Fabian Vehlken\inst{1}\orcidID{0009-0002-1434-3672}}
\and
\anonymize{Thomas Zeume\inst{1}\orcidID{0000-0002-5186-7507}}}
\authorrunning{\anonymize{T. Kneisel, F. Vehlken, T. Zeume}}
\institute{\anonymize{Ruhr University Bochum, Germany}\\
    \email{\anonymize{\{tristan.kneisel,fabian.vehlken,thomas.zeume\}@rub.de}}%
}
\maketitle              %
\begin{abstract}
An important learning objective for computer science students is to learn how to formalize descriptions of real world scenarios in order to subsequently solve real world challenges using methods and algorithms from formal foundations of computer science. Two key steps when formalizing with logical formalisms are to (a) choose a suitable \emph{vocabulary}, that is, e.g.,  which propositional variables or first-order symbols to use, and with which intended meaning, and then to (b) construct actual formal descriptions, i.e.~logical formulas over the chosen vocabulary. While (b) is addressed by several educational support systems for formal foundations of computer science, (a) is so far not addressed at all -- likely because it involves specifying the intended meaning of symbols in natural language.

We propose a conceptual framework for educational tasks where students choose a vocabulary, including an enriched language for describing solution spaces as well as an NLP-approach for checking student attempts and providing feedback. We implement educational tasks for designing propositional and first-order vocabularies within the \Iltis educational system, and report on experiments with data from introductory logic courses for computer science students with > 25.000 data points.   

\end{abstract}

\section{Introduction and Motivation}
\label{section:introduction}

	\begin{figure}[t!]
\begin{subfigure}[t]{0.8\textwidth}
\vspace{-4pt}
  \scalebox{0.6}{%
	\begin{minipage}{0.96\textwidth}
		\begin{tcolorbox}[    tikznode boxed title,
    enhanced,
    arc=0mm,
    interior style={white},
    attach boxed title to top left= {yshift=-\tcboxedtitleheight/2, xshift=1mm},
    fonttitle=\bfseries,
    colbacktitle=white,coltitle=black,
    boxed title style={size=normal,colframe=white,boxrule=0pt},
    title={\hspace{-4mm}Exercise: From modelling to inference (prop. logic)},
    left=4pt,right=4pt,
    ,before upper={\parindent10pt}
		]
			\vspace{2mm}
			\noindent After carefully investigating her faulty software system, Julia has found the following dependencies between the three components of the system:
	\begin{enumerate}
    \item[(a)] If the backend is working correctly, the database is also
working correctly.
    \item[(b)] The backend is only working incorrectly if neither the
database nor the user interface is working correctly.
		\item[(c)] At least one component works correctly.
	\end{enumerate}
	Julia concludes that the database and the backend work correctly. Can you verify her conclusion by modelling the situation in propositional logic and inferring Julia’s conclusion with propositional resolution?

		\end{tcolorbox}
	\end{minipage}
 }
\end{subfigure}%
\begin{subfigure}[t]{0.85\textwidth}
\vspace{-4pt}
\hspace{-40.4mm}%
  \scalebox{0.6}{%
	\begin{minipage}{1.04\textwidth}
		\begin{tcolorbox}[    tikznode boxed title,
    enhanced,
    arc=0mm,
    interior style={white},
    attach boxed title to top left= {yshift=-\tcboxedtitleheight/2, xshift=1mm},
    fonttitle=\bfseries,
    colbacktitle=white,coltitle=black,
    boxed title style={size=normal,colframe=white,boxrule=0pt},
    title={\hspace{-2mm}Exercise: From modelling to inference (first-order logic)},
    left=4pt,right=4pt,
    ,before upper={\parindent10pt}
		]
			\vspace{2mm}
			\noindent
                        Julia is investigating a collection of software packages. Each software package has exactly one maintainer, and software packages may depend on other software packages.
                        She has identified the following dependencies:

			\begin{enumerate}%
                            \item[(1)] No software package is both a program library and a system library.
                            \item[(2)] System libraries depend only on system libraries.
                            \item[(3)] Program libraries may only depend on system libraries which are not maintained by the same maintainer.
                            \item[(4)] Every software package that must be explicitly installed by the user depends on at least one program library directly.
			\end{enumerate}

                        \vspace{-1.3mm}
			\noindent She concludes that there is no system library that must be explicitly installed by the user.
			\noindent Can you confirm her conclusion using methods you learned for first-order logic?
		\end{tcolorbox}
	\end{minipage}

 }
\end{subfigure}
\caption{Typical (simplified) assignments from an introductory course on logic which ask students to formally model real-world scenarios by designing a suitable vocabulary, constructing and manipulating formulas, and inferring conclusions.}
\label{figure:exercises}
\end{figure}

Modelling real-world scenarios with mathematical formalisms and subsequently attacking the underlying problems with rigorous methods is an essential part of computer science education at the university level \cite{ACM2013,GI2016}. This is typically taught in courses on formal foundations of computer science, including courses on logic for computer science.  \cref{figure:exercises} sketches two exercises that students typically can solve after an introductory course on logic by executing the following steps:
\begin{description}
  \item[Step 1:]  Design a suitable \emph{vocabulary} to describe a real world scenario. For instance, for propositional logic, this involves identifying suitable propositional variables and stating their intended meaning. For first-order logic, suitable relation and  function symbols with their intended meaning have to be chosen.  
  \item [Step 2:] Describe the scenario by logical formulas over this vocabulary.
  \item [Step 3:] Transform formulas into an adequate, more simple form.
	\item [Step 4:] Infer new knowledge using an inference mechanism (e.g.\ resolution).%
\end{description}

Such courses on formal foundations are typically among the hardest for computer science students. Improving motivation and understanding in these courses is an important goal, which can be fostered by following the recommendations of the National Research Council of the US, which advocate, among others, to ``Leverage technologies to make the most effective use of students’ time, shifting from information delivery to sense-making and practice in class'' (see \cite{Singer2012,Beach2012}). 

Technological support for Steps 2 -- 4 is present in state-of-the-art systems such as the \Iltis educational system \cite{Iltis,SchmellenkampVZ24}, which also supports combining small educational tasks into assignments where students execute Steps 2 -- 4 working with their own formulas the whole time.

Unfortunately, there is currently no educational support system that supports students in vocabulary design tasks, i.e. there is no support for Step 1. As this is the essential step when translating scenario descriptions given in natural language into formal language, we call this the \emph{natural language gap} in teaching support for modelling with logical formalisms. 

The goal of this paper is to bridge this natural language gap. 

An educational task for designing a suitable vocabulary must allow students to specify (i) which symbols they want to use in their logical formulas -- propositional variables in the case of propositional logic; relation, function, and constant symbols in the case of first-order logic -- and (ii) which meaning the symbols shall have, described in natural language. The system should allow specifications as flexible as possible and still be able to check student attempts and provide feedback, if necessary.

From our perspective, such an educational task needs to
\begin{itemize}[leftmargin=9mm]
 \item[(R1)] offer a simple, intuitive interface to students;
 \item[(R2)] provide immediate feedback with high accuracy; and
 \item[(R3)] be deployable with small resources.
\end{itemize}

Our contributions are as follows:
\begin{itemize}
 \item We propose a conceptual framework for vocabulary design tasks within educational support systems. This includes, among others, (a) a formalism  that allows instructors to specify solution spaces, and (b) approaches for providing feedback using natural language processing and custom algorithms.
 \item We implement such vocabulary design tasks for propositional and first-order logic within the \Iltis educational support system. These tasks can be combined with other tasks to let students execute Steps 1 -- 4 from above in order to solve assignments as sketched in \cref{figure:exercises} with their chosen vocabularies.
 \item We design assignments for an introductory logic course at a German university and evaluate our framework on a data set with > 25.000 data points obtained for these assignments. We discuss variants of our NLP-approach  and how it compares to state-of-the-art LLMs.%
\end{itemize}

\subsubsection{Related work.} There is, to the best of our knowledge, no support in educational support systems for bridging the gap between natural and formal language. A related task is to describe formal constructs in natural language. One such task is provided by AutomataTutor \cite{AntoniKAGV2015,AntoniHKRW2020} which provides natural language feedback on mistakes made by students when designing finite state automata. %

\subsubsection{Outline.} We describe our framework in \cref{section:framework}, discuss implementation aspects in  \cref{section:implementation}, and evaluate our approach in \cref{section:evaluation}.
 
\section{A Framework for Vocabulary Design Tasks}
\label{section:framework}

\begin{figure}
    \begin{minipage}[t]{0.59\linewidth}
        \vspace{0pt}
        \scalebox{0.85}{
            \begin{tikzpicture}
                \node[screenshot] {\includeScreenshot{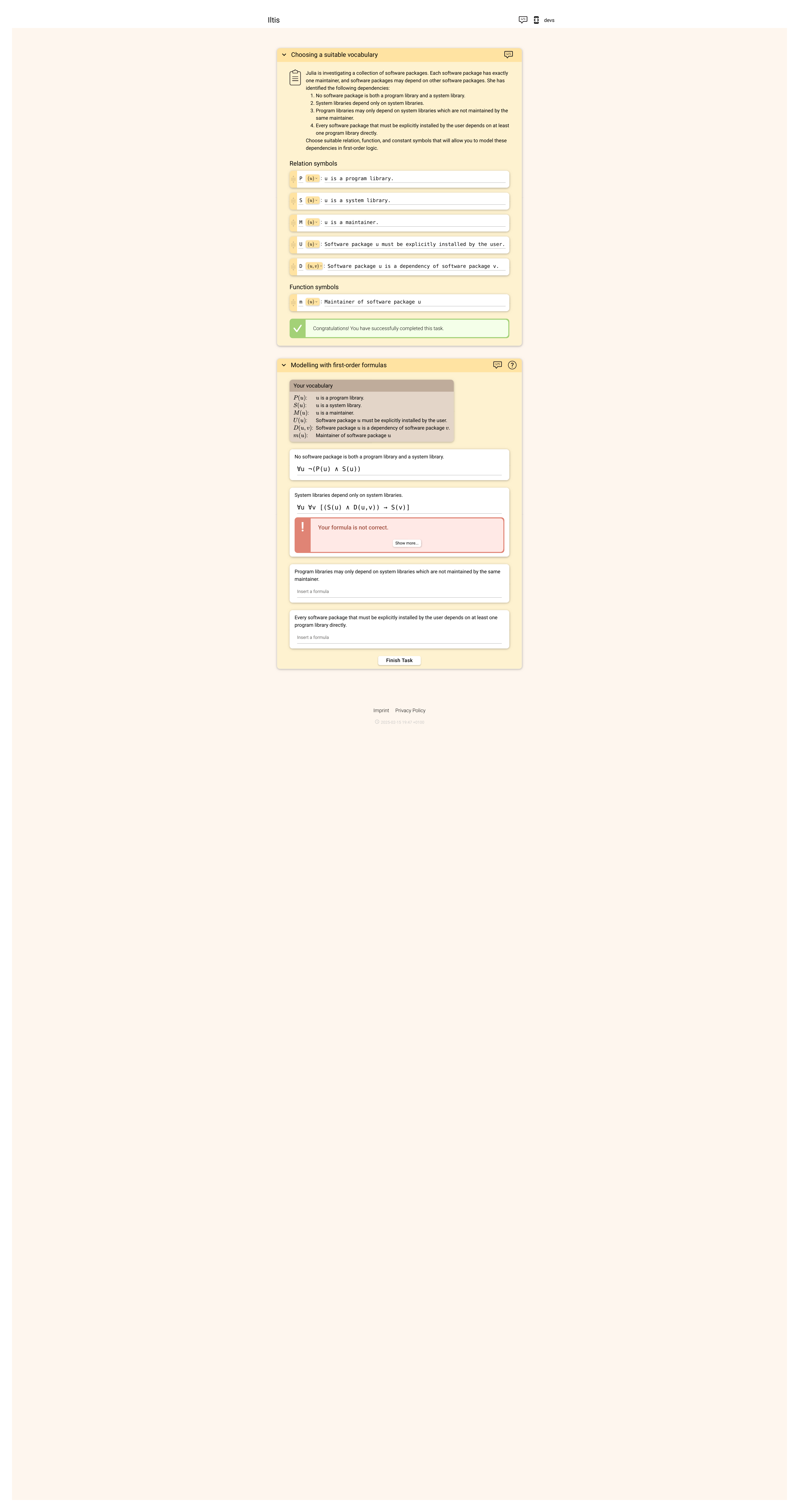}{\linewidth}{293mm}{1235mm}{293mm}{52mm}};
                \node[screenshot border]{\phantom{\includeScreenshot{iltis-screenshots/permuted_second-formula-wrong-wide-morepadding.pdf}{\linewidth}{293mm}{1235mm}{293mm}{52mm}}};
            \end{tikzpicture}
        }
    \end{minipage}
    \hspace{-5mm}
    \begin{minipage}[t]{0.48\linewidth}
        \vspace{0pt}
        \scalebox{0.85}{
            \begin{tikzpicture}
                \node[screenshot] {\includeScreenshot{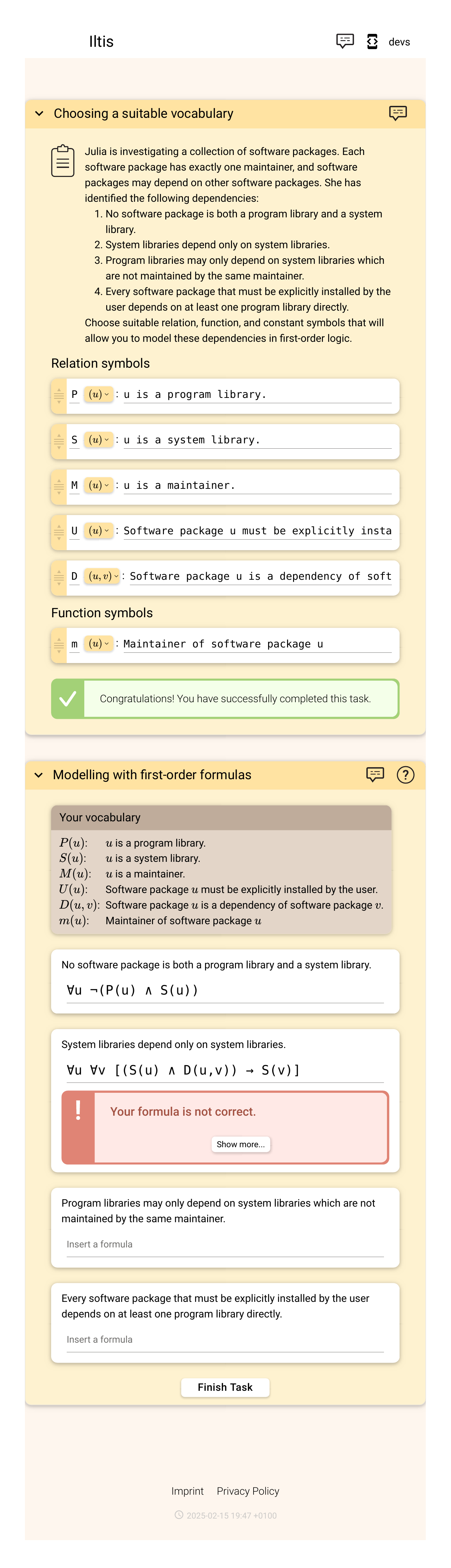}{\linewidth}{13mm}{84mm}{13mm}{389mm}};
                \node[screenshot border]{\phantom{\includeScreenshot{iltis-screenshots/permuted_second-formula-wrong.pdf}{\linewidth}{13mm}{84mm}{13mm}{389mm}}};
            \end{tikzpicture}
        }
    \end{minipage} 
    \caption{An assignment in the \Iltis educational support system where students (a) design a first-order vocabulary (left), and then (b) write first-order formulas using their  vocabulary (right).}
    \label{figure:fotasks}
\end{figure}

When students design a (propositional or first-order) vocabulary with the intention of writing formulas, they must say (i) which symbols they want to include, and (ii) which meaning the symbols shall have. An educational task that allows to specify a vocabulary must then check student attempts for correctness. If an attempt is not correct, it must provide feedback; if it is correct, the vocabulary can subsequently be used in follow-up tasks (compare \cref{figure:fotasks}).

For checking these conditions automatically, an instructor needs to specify (1) which symbols can in principal be used, and (2) which subsets of symbols constitute solutions. More formally, a \emph{solution space} $(\calV, \calS)$ consists of
\begin{itemize}
 \item \emph{potential vocabulary symbols} $v^* = (n^*, D^*) \in \calV$ with a name $n^*$ and a set $D^* = \{d_1^*, d_2^*, \ldots \}$ of natural language descriptions;
 \item a \emph{solution set} $\calS \subseteq 2^\calV$ of subsets of symbols that constitute solutions.
\end{itemize}

\begin{example}\label{ex:potential-vocabulary-symbols}
The solution space for the propositional modelling task from \cref{figure:exercises} could contain potential vocabulary symbols

\begin{itemize}
 \item $(B, \{\text{The backend is correct}, \text{The backend works}, \ldots\})$, and 
 \item $(P, \{\text{The printer works}, \ldots\})$
\end{itemize}

The set $\calS$ could include a set $\{B, D, U\}$ of sufficient propositional variable symbols for backend, database, and user interface working correctly.%

In the first-order modelling task, names of symbols would also include information on whether the symbol is a function, relation or constant symbol and an arity, if applicable, for example 
\begin{itemize}
    \item $ \left(D(u,v), \left\{ 
            \begin{tabular}{l}
                $ \text{Software package } u \text{ depends on software package } v, $\\
                $ \text{Software package } v \text{ is a dependency of software package } u $
            \end{tabular}%
           \right\}\right) $, 
           \vspace{1mm}
    \item $ (m(u), \{ \text{The maintainer of software package } u, \dots \}) $,
\end{itemize}%
where $ D $ is a binary relation symbol and $ m $ is a unary function symbol. Note that the descriptions may refer to parameters depending on the symbol's arity. \qed
\end{example}

One part of our framework will, when given a student attempt $V$ containing \emph{vocabulary symbols} $v = (n, d)$ with a name $n$ and a description $d$, map $v$ to a potential solution symbol $v^* = (n^*, D^*)$ such that $d$ matches some description $d^* \in D^*$ as ``good as possible''. For providing normalized feedback, we measure the quality of a match between $d$ and $d^*$ with categories from a finite set $\calC = \calP \uplus \calN$  of positive categories $\calP$ and negative categories $\calN$. The intuition is that $(d, d^*)$ is assigned a category from $\calP$ if $d$ is evaluated as matching $d^*$ by a human tutor.   

Our framework, when given a student attempt $V$, works in two phases:

\begin{description}
 \item [Phase 1: Mapping vocabulary symbols.]~\\
     \vspace{-3mm}
    \begin{itemize}
	 \item[(a)]  Construct a map $\mu\colon V \rightarrow \calV \times \calC$
        that maps each $v \in V$ to a  pair $(v^*, c)$ where $v^*$ is a potential vocabulary symbol in the solution space that is the best fit for $v$ and $c$ is a category from $\calC = \calP \uplus \calN$ indicating how well $v^*$ fits $v$.
	\item[(b)] If $\mu$ assigns a negative category $c \in \calN$ to some $v$, provide feedback according to the category $c$. If $\mu$ only assigns positive categories from $\calP$, then continue with Phase 2.
    \end{itemize}

  \vspace{2mm}
 \item[Phase 2: Checking solutions.]~\\
     \vspace{-3mm}
    \begin{itemize}
     \item[(a)] Check that the vocabulary symbols $ V $ chosen by the student constitute a solution by testing\footnote{Sets are considered to be multisets to recognize student attempts using multiple symbols for a single symbol of the solution space.} that $\{v^* \mid (v^*, c) \in \mu(V)\} \in \calS$.     
      \item[(b)] If the solution is not correct, provide feedback accordingly. Otherwise, the attempt $V$ with its symbols and descriptions is the output of this educational task (and can be used in subsequent tasks).
 \end{itemize}
\end{description}

In the next two subsections we provide more details on the two phases.

\subsection{Phase 1: Mapping vocabulary symbols}\label{section:mapping-vocabulary-symbols}

Mapping a symbol $v = (n, d)$ from a student attempt into the solution space requires to determine whether, and to which degree, the description $d$ corresponds to natural language descriptions of symbols in the solution space. For descriptions of propositional variables, this is essentially the NLP task of determining the (semantic) similarity of pairs of strings. For first-order vocabulary symbols, this is slightly more involved as descriptions may refer to parameters such as in \cref{ex:potential-vocabulary-symbols}. In particular, the order in which the natural language description refers to parameters can differ, e.g.\ when the description is phrased passively.

We aim for our framework to be easily deployable by instructors with full data sovereignty and few resources. This excludes proprietary LLMs, because data does not remain under the control of instructors, as well as open LLMs since they require too many resources (in terms of hardware but also time/effort per query). We refer to \cref{section:evaluation} for further discussion of this aspect.

For this reason, we started from small similarity models and fine-tuned them with additional data. The necessity for fine-tuning was suggested by initial experiments that indicated that accuracy without fine-tuning is not sufficient, in particular as we required similarity models for the German language (also see \cref{section:evaluation}). To be sensitive to the context of assignments and to avoid the risk of undesirable side-effects, we created a custom model for each of the six assignments in \cref{section:appendix-propositional-exercises,section:appendix-fo-exercises}. 

For fine-tuning for an assignment with solution space $(\calV, \calS)$, we used data sets $\calD$ containing tuples $(d, d^*, c)$ where $d$ is some description, $d^*$ is a description used for some element in $\calV$, and $c$ is a category from $\calC = \calP \uplus \calN$ indicating how well $d$ matches $d^*$. One challenge is that such data cannot be collected from students without an implementation of a vocabulary design task with accurate feedback. To circumvent this, we generated data sets $\calD$ using a grammar-based approach. For each description $d^*$ used in the solution space and each category $c \in \calC$, we designed a context-free grammar that generates descriptions $d$ leading to tuples $(d, d^*, c)$ such that $d$ matches $d^*$ with quality $c$. Using grammars allows to compactly represent many descriptions belonging to the same category. 

\begin{example}\label{ex:best-match-user-interface}
    A grammar for a description $ d^* = $ ``The user interface works correctly'' used in a solution space (cf.\ \cref{figure:exercises}) 
    might look as follows.
    \begin{align*}
        \nonterminal{S} &\to \nonterminal{U}~\nonterminal{W}\\
        \nonterminal{U} &\to \termi{user interface} \mid \termi{user-interface} \mid \termi{UI}\\
        \nonterminal{W} &\to \termi{works}~\nonterminal{C} \mid \termi{is working}~\nonterminal{C} \mid \termi{behaves}~\nonterminal{C} \mid \termi{is behaving}~\nonterminal{C}\\
        \nonterminal{C} &\to \termi{correctly} \mid \termi{properly}%
    \end{align*}
    In addition to $ d^* $ itself, several alternative phrasings are possible, e.g.\ using ``properly'' instead of ``correctly'' or ``UI'' instead of ``user interface''. Descriptions $ d $ derived from the grammar match $ d^* $ with high quality, and tuples $(d, d^*, c)$, for a high category $c$, were included in the training set.\qed
\end{example}

\subsection{Phase 2: Checking solutions}
Suppose that a student attempt $V$ has been mapped to a set $V^*$ of potential vocabulary symbols from the solution space in Phase 1. In Phase 2, our framework checks that $V^*$ constitutes a solution by testing whether $V^* \in \calS$. Additionally, feedback is provided in case $V^* \notin \calS$.

\subsubsection{Specifying and checking solutions.} Specifying $\calS$ can be quite cumbersome as scenarios can often be modelled with many different vocabularies, particularly in the case of first-order vocabularies. In most cases, however, there is a ``canonical vocabulary'' from which all other vocabularies can be derived by small modifications. Examples of such modifications are that (i) constant symbols can also be modelled by unary relation symbols (with the additional constraint that the relation may only contain one element); (ii) $k$-ary function symbols can be modelled by $(k+1)$-ary relation symbols; and (iii) for modelling $k$ types of elements in a domain one can use either $k$ or $k-1$ unary relation symbols.

In order to simplify the specification of the solution set $\calS$, our framework allows to specify proper solutions using Boolean formulas over potential vocabulary symbols. Such a formula, for instance, can state that a constant symbol $c$ or its corresponding unary relation symbol $R_c$ has to be used. For a student attempt $V$ the framework then checks whether it satisfies the Boolean formula. 

\subsubsection{Providing feedback.} The most prominent reason for a student attempt $V$ to be incorrect is that some necessary vocabulary symbol is not included in $V$. To provide feedback in such cases, our framework allows to specify feedback for different sets of potential vocabulary symbols given via Boolean formulas.

For instance, when designing a vocabulary for the propositional assignment in \cref{figure:exercises}, a student may only have specified variables $B$ and $D$ because she only read statement (a). To provide feedback in such cases, an instructor can specify that for student attempts $V$ that satisfy $\neg U$ (i.e.\ $U$ has not been included by the student) the system provides the following feedback 

\begin{itemize}
 \item[]    ``Did you make sure that the statement \emph{The backend is only working incorrectly if neither the database nor the user interface is working correctly.}\\ can be modelled?'' 
 
\end{itemize}

\section{Vocabulary Design Tasks in Practice}
\label{section:implementation}
We implemented vocabulary design tasks for propositional and first-order logic following the framework described in the previous section within the \Iltis educational system. The user interface for students is illustrated in \cref{figure:fotasks}.

The compositional task model of  \Iltis allows instructors to flexibly combine small educational tasks into larger exercises such that student inputs can be used in later tasks. Our vocabulary design tasks complete a task portfolio that enables exercises for propositional and first-order logic where students (1) design a vocabulary, (2) construct formulas for a scenario, (3) transform formulas into normal forms, and (4) infer new knowledge using, for instance, resolution. 

We highlight some aspects of our implementation.

\subsubsection*{Categories for descriptions of vocabulary symbols.} For classifying descriptions of vocabulary symbols we used, for simplicity, a reduced set of categories containing positive categories $\calP = \{\text{C1}, \text{C2},\text{C3}\}$ and negative categories $\calN = \{\text{C4}, \text{C5}\}$. The categories are assigned to pairs $ (d, d^*) $ of descriptions of symbols $ v $ and $ v^* $ according to the following criteria:
\begin{description}
 \item [C1: ] The description $d$ of $v$ is synonymous to description $d^*$ of $v^*$.
 \item [C2: ] The description $d$ of $v$ matches the description $d^*$ of $v^*$ almost entirely, except for minor issues such as missing unimportant words.
 \item [C3: ] The description $d$ contains a unique reference to $d^*$, even though it is not a proper description.
 \item [C4: ] The description $d$ of $v$ does not match description $d^*$ of $v^*$, but is vaguely related to it.
 \item [C5: ] The descriptions $d$ of $v$ and $d^*$ of $v^*$ are unrelated.
\end{description}

These categories provide a preliminary scale of correctness and allow to provide feedback. As an example, for the scenario from \cref{figure:exercises}, a useful propositional variable is $D$ with the intended meaning ``The database works correctly''. Taking this to be $ d^* $, examples of a description $ d $ for each of the categories are: C1 ``The database runs properly''; C2 ``Database correct''; C3 ``Database''; C4 ``Something is correct''; C5 ``The sun is shining''.

\subsubsection*{Enabling workflows for first-order logic} Providing flexibility in how students can design their vocabulary comes with the following challenge. In many assignments, we ask students to write formulas using the vocabulary they have designed in a previous step. Solutions for such a task may be specified in some vocabulary $V^*$, while a student has designed vocabulary $V$.  To support this, our implementation supports translations of formulas over some vocabulary $V^*$ into formulas over vocabulary $V$. This allows students to continue, for example, with a vocabulary with a constant symbol $c$ or a corresponding unary relation $R_c$.%
\subsubsection*{Use in the classroom} 
We used web-based assignments that included the educational tasks for designing vocabularies for propositional and first-order logic in an introductory course on logic at \anonymize{Ruhr University Bochum} over two semesters in 2023 and 2024. The course introduces students to modelling, normal forms, and inference in propositional logic, modal logic, and first-order logic. 

We created four assignments for propositional logic and two assignments for first-order logic in different modelling contexts, including CS scenarios and non-CS scenarios (see \cref{section:appendix-propositional-exercises,section:appendix-fo-exercises}). In all of them, students are asked to design a vocabulary which is then used for follow-up tasks where students write formulas etc. (see also \cref{table:assignment-overview} for an overview). As the course language is German, all assignments and NLP models are specifically for German.
 
\section{Experimental Evaluation}\label{sec:experiments}
\label{section:evaluation}
In this section, we evaluate the accuracy of NLP-based methods for classifying student attempts in vocabulary design tasks. Our focus is on evaluating Phase~1 of our approach, that is, mapping student descriptions into the solution space. In particular, we address the following questions:
\begin{itemize}
 \item With which accuracy does automatic classification of student-provided descriptions agree with a human tutor classification?
 \item How does the accuracy of small fine-tuned models for semantic similarity compare to the accuracy of large language models?
\end{itemize}

To answer these questions, we compare six NLP methods. Five of them are based on a small semantic similarity NLP model; the last one is based on an LLM. For each of the methods, we evaluate its performance for (a) binary classification into positive categories $\calP$ and negative categories $\calN$, and (b) multi-class classification into the finer categories C1 -- C5. We test their performance for vocabulary design tasks for propositional and first-order vocabularies. 

We first sketch the set-up and methods, then we report on the performance.

\subsection{Data collection and Preparation} \label{section:data}

We collected authentic student inputs for vocabulary design tasks in the introductory course on logic described in \cref{section:implementation}. For all assignments, we collected student attempts $v = (n, d)$ with a name $n$ and description~$d$, see \cref{table:assignment-overview} for an overview. We refer to the appendix for more details on the assignments.

\begin{table}[t]
    \newcommand{\slimcol}{\hspace{-8pt}}
    \setlength\tabcolsep{0.9mm}
    \begin{tabularx}{\textwidth}{l|X<{\raggedright}|c<{}|c<{}|c<{}|c<{}}
        \toprule id & context & size of $\calV$ & \# inputs & \# distinct inputs & \# grammar sample \\
        \midrule
        \multicolumn{6}{c}{\textbf{Propositional logic}}\\
        \midrule
        P1 & Lecture participation & 3 & 4156 & 624 & 2040 \\
				P2 & Movie night & 4 & 4132 & 711 & 1724\\
				P3 & Chat system & 4 & 1458 & 430 & 7464\\
				P4 & Hardware failures & 5 & 1084 & 305 & 2440 \\
        \midrule
        \multicolumn{2}{c|}{$\sum$} & & 10830 & 2070 &  \\
        \midrule
        \multicolumn{6}{c}{\textbf{First-order logic}}\\
        \midrule
        F1 & Book collection & 9 & 11534 & 2413 & 2299 \\
        F2 & Faculty conference & 8 & 6144 & 1655 & 2262 \\
        \midrule
        \multicolumn{2}{c|}{$\sum$} & & 17678 & 4068 &  \\
        \midrule
    \end{tabularx}
    \caption{Assignments used in the evaluation with size $|\calV|$ of the solution space (i.e. number of potential vocabulary symbols), number of student attempts, number of normalized distinct attempts, and size of grammar-generated training set. 
    }\label{table:assignment-overview}
\end{table}

The data was used for (1) fine-tuning NLP models and (2) evaluating different NLP methods. To this end, descriptions entered by students were normalized by (i) replacing all letters by their lower-case variants, (ii) removing superfluous white space, and (iii) removing some special characters (such as quotation marks). This is motivated by the assumption that students do not consider \mbox{(i) -- (iii)} as important aspects when describing vocabulary symbols.

We then hand-labelled all pairs $(d, d^*)$ of normalized descriptions $d$ and descriptions $d^*$ occurring in the solution space with a category $c_h(d, d^*)$ from \mbox{C1 -- C5}. This hand-labelled data was split into blocks of 80\% for fine-tuning NLP models and 20\% for evaluation.  More precisely, for the propositional logic assignments, 2142 student descriptions  were retained for evaluation, corresponding to 7929 pairs (of one of the student descriptions and one description in the solution space). Of these, 1879 were ``positive'' pairs hand-labelled with a category in $\calP$ and 6050 were ``negative'' pairs labelled with a category in $\calN$. For the first-order logic exercises, 3367 student descriptions were retained for evaluation, corresponding to 29144 pairs (2342 positive, 26802 negative).

\subsection{NLP-based Methods} 

We evaluate six NLP-based methods: %

\begin{description}
    \item[(1) \texttt{Bert-Base}:] The semantic similarity German BERT model \cite{BERT-base-german-uncased25}. (This model is the basis for fine-tuning models for methods (2) -- (5).)
    \item[(2) \texttt{Bert+Grammar+Per-assignment}:] One fine-tuned BERT model (with (1) as base) for each assignment with data generated with grammars according to the approach described in \cref{section:mapping-vocabulary-symbols}. (This model was used to provide feedback for student attempts during data collection.)
     \item[(3) \texttt{Bert+Grammar+Single-model}:] Like (2), but only a single fine-tuned BERT model for all assignments (but one model per logic), i.e.\ with data generated with the grammars of all the assignments.
     \item[(4) \texttt{Bert+Student+Per-assignment}:] One BERT model (from (1)) fine-tuned for each assignment with human-labelled data (see \cref{section:data}).
     
     \item[(5) \texttt{Bert+Student+Single-model}:] Like (4), but only a single fine-tuned BERT model for all assignments (but one model per logic).
     
    \item[(6) \texttt{gpt}:] Query to gpt-4o-mini with an assignment specific prompt containing the description text of the exercise and multiple examples for classifications. %
\end{description}

For inference, the models of methods (1) -- (5) are deployed on a server with an NVIDIA RTX 4090.
They are all very small ($ \approx 111 $ million parameters) and take $ \approx 0.005 $ seconds on average to answer a query. Querying gpt-4o-mini via API with our assignment-specific prompts takes $ \approx 0.9$ seconds on average.\footnote{OpenAI has not revealed the size of gpt-4o-mini. They only compare its performance to other models, whose sizes are also unknown, see \cite{GPT-4o-mini-announcement}.}

Only the methods (2), (3), and (6) can be used out-of-the-box in an educational support system, as the other methods (1), (4), and (5) require to fix thresholds for classifying inputs into categories. More precisely, the NLP models (1) -- (5) assign numerical values to pairs. For classifying inputs into categories, thresholds for these values need to be specified. For (2) and (3), these thresholds are fixed when fine-tuning the base Bert model with data specifically generated with grammars. For (1), (4), and (5), good thresholds can only be guessed or determined after looking at actual data (also see evaluation methodology below).

\subsection{Evaluation Methodology} 

We compare the NLP-based methods with respect to their performance in (a) binary classification into positive categories $\calP$ and negative categories $\calN$, and (b) multi-class classification into the finer categories C1 -- C5. 
To this end, for (a), we use each of the methods to classify each pair $(d, d^*)$ of a description $d$ provided by a student and description $d^*$ from a solution space. If the result is consistent with whether the hand-labelled category $c_h(d, d^*)$ is in $\calP$ or $\calN$, then the pair is counted as being correctly classified. 
For comparing the methods, we compare the ratios of correctly classified pairs, ratios of false positives and false negatives, and the ratio \condPos\ of correctly classified pairs whose hand-labelled category is in $\calP$, likewise  \condNeg\ for $\calN$. For (b), we proceed similarly but let the methods classify each pair $(d, d^*)$ into one of the categories C1 -- C5 and compare the result with the hand-labelled category $c_h(d, d^*)$. %

As described above, methods (1), (4), and (5) require to fix thresholds for  numerical values in order to assign categories to inputs. For the evaluation, we chose to make these methods ``as good as possible'' by picking the thresholds such that the ratio of correctly classified pairs is maximized.\footnote{The Nelder-Mead method provided by the \texttt{scipy.optimize} Python module is used.}

\begin{table}
\begin{subtable}[h]{\textwidth}
    \setlength\tabcolsep{3.5mm}
    \begin{tabularx}{\textwidth}{l|r<{}|r<{}|r<{}|r<{}|r<{}}
        \toprule method & \multicolumn{1}{c|}{correct} & false-pos. & false-neg. & $ \condPos $ & $ \condNeg $ \\
        \midrule
        \multicolumn{6}{c}{\textbf{Propositional logic}}\\
        \midrule
        \methodB   & 92.90\% & 0.48\% & 6.62\% & 72.06\% & 99.37\% \\
        \methodBGP & 98.41\% & 0.79\% & 0.79\% & 96.65\% & 98.96\% \\
        \methodBGO & 99.00\% & 0.28\% & 0.72\% & 96.97\% & 99.64\% \\
        \methodBSP & 99.95\% & 0.03\% & 0.03\% & 99.89\% & 99.97\% \\
        \methodBSO & 99.76\% & 0.04\% & 0.20\% & 99.15\% & 99.95\% \\
        \methodGPT & 99.00\% & 0.47\% & 0.53\% & 97.76\% & 99.39\% \\
       \midrule
        \multicolumn{6}{c}{\textbf{First-order logic}}\\
        \midrule
        \methodB   & 94.07\% & 2.74\% & 3.19\% & 60.33\% & 80.55\% \\
        \methodBGP & 96.03\% & 1.91\% & 2.06\% & 74.34\% & 81.45\% \\
        \methodBGO & 95.88\% & 2.29\% & 1.84\% & 77.16\% & 81.03\% \\
        \methodBSP & 98.86\% & 0.44\% & 0.70\% & 91.29\% & 83.05\% \\
        \methodBSO & 98.90\% & 0.32\% & 0.79\% & 90.18\% & 83.18\% \\
        \methodGPT & 94.38\% & 3.40\% & 2.22\% & 72.37\% & 79.83\% \\
        \midrule
    \end{tabularx}
    \subcaption{Aggregated accuracy of NLP-based methods in binary classification.}%
	 \label{table:binary}
\end{subtable}

   \begin{subtable}[h]{\textwidth}
   \setlength\tabcolsep{3.5mm}
    \begin{tabularx}{\textwidth}{l|r<{}|r<{}|r<{}|r<{}|r<{}}
        \toprule method & \multicolumn{1}{c|}{correct}  & too-high & too-low & $ \condPos $ & $ \condNeg $ \\
        \midrule
        \multicolumn{6}{c}{\textbf{Propositional logic}}\\
        \midrule
        \methodB   & 6.87\%  & 70.32\% & 22.80\% & 80.63\% & 46.00\%  \\
        \methodBGP & 72.10\% & 1.55\%  & 26.35\% & 89.41\% & 99.07\%  \\
        \methodBGO & 71.55\% & 2.18\%  & 26.27\% & 89.84\% & 99.75\%  \\
        \methodBSP & 77.36\% & 0.01\%  & 22.63\% & 87.01\% & 100.00\% \\
        \methodBSO & 76.50\% & 0.29\%  & 23.21\% & 88.19\% & 99.97\%  \\
        \methodGPT & 73.57\% & 0.43\%  & 26.01\% & 78.07\% & 99.82\%  \\
        \midrule
        \multicolumn{6}{c}{\textbf{First-order logic}}\\
        \midrule
        \methodB   & 18.42\% & 74.14\% & 7.45\% & 61.91\% & 79.86\% \\
        \methodBGP & 85.28\% & 6.79\%  & 7.93\% & 68.87\% & 82.27\% \\
        \methodBGO & 86.06\% & 5.84\%  & 8.10\% & 72.54\% & 81.92\% \\
        \methodBSP & 90.41\% & 0.61\%  & 8.98\% & 76.77\% & 83.40\% \\
        \methodBSO & 90.86\% & 0.80\%  & 8.34\% & 79.97\% & 83.35\% \\
        \methodGPT & 87.03\% & 4.20\%  & 8.77\% & 66.61\% & 80.04\% \\
        \midrule
    \end{tabularx}
    \subcaption{Aggregated accuracy of NLP-based methods in multi-class classification.}\label{table:multi-class}	    
   \end{subtable}
\caption{Accuracy of NLP-based methods in (a) binary classification and (b) multi-class classification on authentic student data, measured in ratio of correctly classified pairs, ratio of false positives and false negatives, and ratios of correctly classified positive pairs and correctly classified negative pairs.}
\label{table:evaluation}
\end{table}

\subsection{Evaluation and Discussion}

The accumulated evaluation results for binary and multi-class classification are summarized in \cref{table:evaluation}. Detailed results, broken down according to individual assignments, are provided in the appendix. The evaluation was performed on the 20\% of student descriptions retained for evaluation, corresponding to 7929 pairs $(d, d^*)$ for propositional logic and 29144 pairs for first-order logic.

For binary classification, all methods perform with $> 90 \%$ accuracy for propositional and first-order logic. For propositional logic, all methods but \texttt{Bert-Base} reach $> 98 \%$ accuracy. For first-order logic $> 98 \%$ accuracy is only reached by models fine-tuned with authentic student data; with all other but \texttt{Bert-Base} having $~ 95 \%$ accuracy. For multi-class classification, all methods perform worse. All methods except \texttt{Bert-Base} reach $> 70 \%$ accuracy. Again, models fine-tuned with authentic student data perform best. Interestingly, the models for first-order vocabularies perform much better with $> 10 \%$ better accuracy than for propositional vocabularies.

In summary, the accuracy of methods (2) -- (5) is very promising. In the context of educational support systems, the LLM based method (6) has the drawback that it requires much more resources (in terms of time and energy) and, in case of proprietary models, also has consequences for data sovereignty.  Using methods (2) or (3) when no authentic student data for fine-tuning is available and fine-tuning with authentic hand-labelled data as in (4) and (5) when such data becomes available seems to be a good strategy at the moment. 

We expect that the small differences in accuracy between small fine-tuned models and LLMs will disappear soon. It is less clear whether it will be possible in the near future to distil such LLMs into small models in order to save resources.
 
\section{Summary and Perspectives}
\label{section:discussion}
We have presented a framework and an implementation for propositional and first-order vocabulary design tasks that bridges the natural language gap for logical modelling in educational support systems. Our evaluation shows that NLP-based methods are sufficiently accurate for use in an educational support system. We have shown that grammar-generated data can compensate for an initial lack of authentic student data. In particular, using the NLP-based methods (2) -- (5), our system satisfies requirements (R1) --  (R3) stated in the introduction. 

In future work we plan to study how natural language gaps in other subfields of formal foundations of computer science can be bridged.  For instance, we aim for educational tasks for explaining formal languages -- e.g. described by finite state machines, context-free grammars and pushdown automata, as well as Turing machines -- in natural language. 
 
\begin{credits}
\subsubsection{\ackname}

We thank \anonymize{Timo Hönig and Bilal Zafar} for insightful discussions on aspects of LLMs.

This work was supported by \anonymize{the Deutsche Forschungsgemeinschaft (DFG, German Research Foundation), grants 448468041 and 532727578}.

\subsubsection{\discintname}
The authors have no competing interests to declare that are
relevant to the content of this article. 
\end{credits}

\bibliographystyle{splncs04}
\bibliography{bibliography}

\begin{thebibliography}{10}
\providecommand{\url}[1]{\texttt{#1}}
\providecommand{\urlprefix}{URL }
\providecommand{\doi}[1]{https://doi.org/#1}

\bibitem{BERT-base-german-uncased25}
{Bayerische Staatsbibliothek}: bert-base-german-uncased (revision b705f0e)
  (2025). \doi{10.57967/hf/4378},
  \url{https://huggingface.co/dbmdz/bert-base-german-uncased}

\bibitem{Beach2012}
Beach, A.L., Henderson, C., Finkelstein, N.: Facilitating change in
  undergraduate {STEM} education. Change: The Magazine of Higher Learning
  \textbf{44}(6),  52--59 (2012)

\bibitem{AntoniHKRW2020}
D'Antoni, L., Helfrich, M., K\v{r}et\'{i}nsk\'{y}, J., Ramneantu, E.,
  Weininger, M.: Automata tutor v3. In: Lahiri, S.K., Wang, C. (eds.) Computer
  Aided Verification -- 32nd International Conference, {CAV} 2020, Proceedings,
  Part {II}. Lecture Notes in Computer Science, vol. 12225, pp. 3--14. Springer
  (2020). \doi{10.1007/978-3-030-53291-8\_1}

\bibitem{AntoniKAGV2015}
D'Antoni, L., Kini, D., Alur, R., Gulwani, S., Viswanathan, M., Hartmann, B.:
  How can automatic feedback help students construct automata? {ACM}
  Transactions on Computer-Human Interaction  \textbf{22}(2),  pp.~9:1--9:24
  (2015). \doi{10.1145/2723163}

\bibitem{GI2016}
{Gesellschaft für Informatik e.\,V.}: Empfehlungen für {B}achelor- und
  {M}aster-{P}rogramme im {S}tudienfach {I}nformatik an {H}ochschulen.
  \url{https://gi.de} (2016), \url{https://dl.gi.de/handle/20.500.12116/2351}

\bibitem{Iltis}
Iltis: Formal foundations of computer science online,
  \url{https://iltis.rub.de}

\bibitem{ACM2013}
{Joint Task Force on Computing Curricula, Association for Computing Machinery
  (ACM)}, {IEEE Computer Society}: Computer Science Curricula 2013: Curriculum
  Guidelines for Undergraduate Degree Programs in Computer Science. Association
  for Computing Machinery, New York, NY, USA (2013)

\bibitem{GPT-4o-mini-announcement}
{OpenAI}: Gpt-4o mini: advancing cost-efficient intelligence.
  \url{https://openai.com/index/gpt-4o-mini-advancing-cost-efficient-intelligence/}
  (2024), accessed: 15 Feb 2025

\bibitem{SchmellenkampVZ24}
Schmellenkamp, M., Vehlken, F., Zeume, T.: Teaching formal foundations of
  computer science with {Iltis}. Educational Column of the Bulletin of {EATCS}
  (2024),
  \url{http://bulletin.eatcs.org/index.php/beatcs/article/download/797/842}

\bibitem{Singer2012}
Singer, S.R., Nielsen, N.R., Schweingruber, H.A.: Discipline-based education
  research. Washington, DC: The National Academies  (2012)

\end{thebibliography}

\newpage
\appendix

\section{Detailed Evaluation Data}

The detailed evaluation results for binary and multi-class classification are
presented in Tables \cref{table:binary-evaluation} and \cref{table:multi-class-evaluation}.

\begin{table}
    \scriptsize
    \begin{tabularx}{\textwidth}{l|X<{\raggedright}|r<{}|r<{}|r<{}|r<{}|r<{}|r<{}|r<{}|r<{}|r<{}|r<{}}
        \toprule method & \multicolumn{1}{c|}{id} & \multicolumn{1}{c|}{$\noComparisons$} & \#inputs & \#pairs & \#pos & \#neg & \multicolumn{1}{c|}{correct}  & false-pos. & false-neg. & $ \condPos $ & $ \condNeg $ \\
        \midrule
        \multicolumn{12}{c}{\textbf{Propositional logic}}\\
        \midrule
        \methodB   & P1           & 3 & 846  & 2538 & 737  & 1801 & 91.96\%  & 0.00\% & 8.04\% & 72.32\%  & 100.00\% \\
        \methodB   & P2           & 4 & 805  & 3220 & 678  & 2542 & 95.12\%  & 0.00\% & 4.88\% & 76.84\%  & 100.00\% \\
        \methodB   & P3           & 4 & 284  & 1136 & 269  & 867  & 90.76\%  & 0.00\% & 9.24\% & 60.97\%  & 100.00\% \\
        \methodB   & P4           & 5 & 207  & 1035 & 195  & 840  & 90.63\%  & 3.67\% & 5.70\% & 69.74\%  & 95.48\%  \\
        \methodB   & \accumulated &   & 2142 & 7929 & 1879 & 6050 & 92.90\%  & 0.48\% & 6.62\% & 72.06\%  & 99.37\%  \\
        \methodBGP & P1           & 3 & 846  & 2538 & 737  & 1801 & 99.29\%  & 0.12\% & 0.59\% & 97.96\%  & 99.83\%  \\
        \methodBGP & P2           & 4 & 805  & 3220 & 678  & 2542 & 97.73\%  & 1.30\% & 0.96\% & 95.43\%  & 98.35\%  \\
        \methodBGP & P3           & 4 & 284  & 1136 & 269  & 867  & 97.36\%  & 1.41\% & 1.23\% & 94.80\%  & 98.15\%  \\
        \methodBGP & P4           & 5 & 207  & 1035 & 195  & 840  & 99.52\%  & 0.19\% & 0.29\% & 98.46\%  & 99.76\%  \\
        \methodBGP & \accumulated &   & 2142 & 7929 & 1879 & 6050 & 98.41\%  & 0.79\% & 0.79\% & 96.65\%  & 98.96\%  \\
        \methodBGO & P1           & 3 & 846  & 2538 & 737  & 1801 & 99.21\%  & 0.04\% & 0.75\% & 97.42\%  & 99.94\%  \\
        \methodBGO & P2           & 4 & 805  & 3220 & 678  & 2542 & 98.82\%  & 0.56\% & 0.62\% & 97.05\%  & 99.29\%  \\
        \methodBGO & P3           & 4 & 284  & 1136 & 269  & 867  & 98.50\%  & 0.18\% & 1.32\% & 94.42\%  & 99.77\%  \\
        \methodBGO & P4           & 5 & 207  & 1035 & 195  & 840  & 99.61\%  & 0.10\% & 0.29\% & 98.46\%  & 99.88\%  \\
        \methodBGO & \accumulated &   & 2142 & 7929 & 1879 & 6050 & 99.00\%  & 0.28\% & 0.72\% & 96.97\%  & 99.64\%  \\
        \methodBSP & P1           & 3 & 846  & 2538 & 737  & 1801 & 99.88\%  & 0.04\% & 0.08\% & 99.73\%  & 99.94\%  \\
        \methodBSP & P2           & 4 & 805  & 3220 & 678  & 2542 & 99.97\%  & 0.03\% & 0.00\% & 100.00\% & 99.96\%  \\
        \methodBSP & P3           & 4 & 284  & 1136 & 269  & 867  & 100.00\% & 0.00\% & 0.00\% & 100.00\% & 100.00\% \\
        \methodBSP & P4           & 5 & 207  & 1035 & 195  & 840  & 100.00\% & 0.00\% & 0.00\% & 100.00\% & 100.00\% \\
        \methodBSP & \accumulated &   & 2142 & 7929 & 1879 & 6050 & 99.95\%  & 0.03\% & 0.03\% & 99.89\%  & 99.97\%  \\
        \methodBSO & P1           & 3 & 846  & 2538 & 737  & 1801 & 99.96\%  & 0.00\% & 0.04\% & 99.86\%  & 100.00\% \\
        \methodBSO & P2           & 4 & 805  & 3220 & 678  & 2542 & 99.60\%  & 0.09\% & 0.31\% & 98.53\%  & 99.88\%  \\
        \methodBSO & P3           & 4 & 284  & 1136 & 269  & 867  & 99.91\%  & 0.00\% & 0.09\% & 99.63\%  & 100.00\% \\
        \methodBSO & P4           & 5 & 207  & 1035 & 195  & 840  & 99.61\%  & 0.00\% & 0.39\% & 97.95\%  & 100.00\% \\
        \methodBSO & \accumulated &   & 2142 & 7929 & 1879 & 6050 & 99.76\%  & 0.04\% & 0.20\% & 99.15\%  & 99.95\%  \\
        \methodGPT & P1           & 3 & 846  & 2538 & 737  & 1801 & 98.50\%  & 0.51\% & 0.99\% & 96.61\%  & 99.28\%  \\
        \methodGPT & P2           & 4 & 805  & 3220 & 678  & 2542 & 99.35\%  & 0.50\% & 0.16\% & 99.26\%  & 99.37\%  \\
        \methodGPT & P3           & 4 & 284  & 1136 & 269  & 867  & 98.68\%  & 0.53\% & 0.79\% & 96.65\%  & 99.31\%  \\
        \methodGPT & P4           & 5 & 207  & 1035 & 195  & 840  & 99.52\%  & 0.19\% & 0.29\% & 98.46\%  & 99.76\%  \\
        \methodGPT & \accumulated &   & 2142 & 7929 & 1879 & 6050 & 99.00\%  & 0.47\% & 0.53\% & 97.76\%  & 99.39\%  \\
        \midrule
        \multicolumn{12}{c}{\textbf{First-order logic}}\\
        \midrule
        \methodB   & F1           & 9 & 2208 & 19872 & 1426 & 18446 & 96.86\% & 0.24\% & 2.90\% & 59.54\% & 75.81\% \\
        \methodB   & F2           & 8 & 1159 & 9272  & 916  & 8356  & 88.10\% & 8.10\% & 3.80\% & 61.57\% & 91.01\% \\
        \methodB   & \accumulated &   & 3367 & 29144 & 2342 & 26802 & 94.07\% & 2.74\% & 3.19\% & 60.33\% & 80.55\% \\
        \methodBGP & F1           & 9 & 2208 & 19872 & 1426 & 18446 & 96.96\% & 1.68\% & 1.37\% & 80.93\% & 74.25\% \\
        \methodBGP & F2           & 8 & 1159 & 9272  & 916  & 8356  & 94.05\% & 2.41\% & 3.55\% & 64.08\% & 97.33\% \\
        \methodBGP & \accumulated &   & 3367 & 29144 & 2342 & 26802 & 96.03\% & 1.91\% & 2.06\% & 74.34\% & 81.45\% \\
        \methodBGO & F1           & 9 & 2208 & 19872 & 1426 & 18446 & 96.72\% & 1.91\% & 1.36\% & 81.00\% & 74.00\% \\
        \methodBGO & F2           & 8 & 1159 & 9272  & 916  & 8356  & 94.06\% & 3.10\% & 2.85\% & 71.18\% & 96.57\% \\
        \methodBGO & \accumulated &   & 3367 & 29144 & 2342 & 26802 & 95.88\% & 2.29\% & 1.84\% & 77.16\% & 81.03\% \\
        \methodBSP & F1           & 9 & 2208 & 19872 & 1426 & 18446 & 99.27\% & 0.30\% & 0.43\% & 93.97\% & 75.74\% \\
        \methodBSP & F2           & 8 & 1159 & 9272  & 916  & 8356  & 97.98\% & 0.74\% & 1.27\% & 87.12\% & 99.17\% \\
        \methodBSP & \accumulated &   & 3367 & 29144 & 2342 & 26802 & 98.86\% & 0.44\% & 0.70\% & 91.29\% & 83.05\% \\
        \methodBSO & F1           & 9 & 2208 & 19872 & 1426 & 18446 & 99.30\% & 0.21\% & 0.49\% & 93.20\% & 75.83\% \\
        \methodBSO & F2           & 8 & 1159 & 9272  & 916  & 8356  & 98.03\% & 0.54\% & 1.43\% & 85.48\% & 99.40\% \\
        \methodBSO & \accumulated &   & 3367 & 29144 & 2342 & 26802 & 98.90\% & 0.32\% & 0.79\% & 90.18\% & 83.18\% \\
        \methodGPT & F1           & 9 & 2208 & 19872 & 1426 & 18446 & 94.50\% & 4.04\% & 1.46\% & 79.66\% & 71.71\% \\
        \methodGPT & F2           & 8 & 1159 & 9272  & 916  & 8356  & 94.11\% & 2.04\% & 3.85\% & 61.03\% & 97.74\% \\
        \methodGPT & \accumulated &   & 3367 & 29144 & 2342 & 26802 & 94.38\% & 3.40\% & 2.22\% & 72.37\% & 79.83\% \\
        \midrule
    \end{tabularx}
    \caption{Accuracy of NLP-based methods in \emph{binary classification}.
        Each method is evaluated individually for each assignment, which is indicated in the id column.
        Id \accumulated is used for the aggregated values that can also be found in \cref{table:binary}.
    }\label{table:binary-evaluation}
\end{table}

\begin{table}
    \scriptsize
    \begin{tabularx}{\textwidth}{l|X<{\raggedright}|r<{}|r<{}|r<{}|r<{}|r<{}|r<{}|r<{}|r<{}|r<{}|r<{}}
        \toprule method & \multicolumn{1}{c|}{id} & \multicolumn{1}{c|}{$\noComparisons$} & \#inputs & \#pairs & \#pos & \#neg & \multicolumn{1}{c|}{correct}  & too-high & too-low & $ \condPos $ & $ \condNeg $ \\
        \midrule
        \multicolumn{12}{c}{\textbf{Propositional logic}}\\
        \midrule
        \methodB   & P1           & 3 & 846  & 2538 & 737  & 1801 & 4.14\%  & 67.26\% & 28.61\% & 85.48\% & 25.49\%  \\
        \methodB   & P2           & 4 & 805  & 3220 & 678  & 2542 & 11.21\% & 68.45\% & 20.34\% & 87.02\% & 47.84\%  \\
        \methodB   & P3           & 4 & 284  & 1136 & 269  & 867  & 4.84\%  & 73.59\% & 21.57\% & 82.90\% & 32.53\%  \\
        \methodB   & P4           & 5 & 207  & 1035 & 195  & 840  & 2.32\%  & 80.10\% & 17.58\% & 36.92\% & 98.33\%  \\
        \methodB   & \accumulated &   & 2142 & 7929 & 1879 & 6050 & 6.87\%  & 70.32\% & 22.80\% & 80.63\% & 46.00\%  \\
        \methodBGP & P1           & 3 & 846  & 2538 & 737  & 1801 & 64.62\% & 2.21\%  & 33.18\% & 86.97\% & 99.94\%  \\
        \methodBGP & P2           & 4 & 805  & 3220 & 678  & 2542 & 70.47\% & 1.52\%  & 28.01\% & 89.09\% & 98.51\%  \\
        \methodBGP & P3           & 4 & 284  & 1136 & 269  & 867  & 75.18\% & 1.41\%  & 23.42\% & 90.33\% & 98.27\%  \\
        \methodBGP & P4           & 5 & 207  & 1035 & 195  & 840  & 92.17\% & 0.19\%  & 7.63\%  & 98.46\% & 99.76\%  \\
        \methodBGP & \accumulated &   & 2142 & 7929 & 1879 & 6050 & 72.10\% & 1.55\%  & 26.35\% & 89.41\% & 99.07\%  \\
        \methodBGO & P1           & 3 & 846  & 2538 & 737  & 1801 & 63.08\% & 3.62\%  & 33.29\% & 86.30\% & 99.94\%  \\
        \methodBGO & P2           & 4 & 805  & 3220 & 678  & 2542 & 70.78\% & 1.40\%  & 27.83\% & 91.89\% & 99.49\%  \\
        \methodBGO & P3           & 4 & 284  & 1136 & 269  & 867  & 73.86\% & 2.64\%  & 23.50\% & 90.33\% & 100.00\% \\
        \methodBGO & P4           & 5 & 207  & 1035 & 195  & 840  & 92.17\% & 0.58\%  & 7.25\%  & 95.38\% & 99.88\%  \\
        \methodBGO & \accumulated &   & 2142 & 7929 & 1879 & 6050 & 71.55\% & 2.18\%  & 26.27\% & 89.84\% & 99.75\%  \\
        \methodBSP & P1           & 3 & 846  & 2538 & 737  & 1801 & 74.23\% & 0.04\%  & 25.73\% & 96.88\% & 100.00\% \\
        \methodBSP & P2           & 4 & 805  & 3220 & 678  & 2542 & 79.10\% & 0.00\%  & 20.90\% & 92.33\% & 100.00\% \\
        \methodBSP & P3           & 4 & 284  & 1136 & 269  & 867  & 76.50\% & 0.00\%  & 23.50\% & 82.90\% & 100.00\% \\
        \methodBSP & P4           & 5 & 207  & 1035 & 195  & 840  & 80.58\% & 0.00\%  & 19.42\% & 36.92\% & 100.00\% \\
        \methodBSP & \accumulated &   & 2142 & 7929 & 1879 & 6050 & 77.36\% & 0.01\%  & 22.63\% & 87.01\% & 100.00\% \\
        \methodBSO & P1           & 3 & 846  & 2538 & 737  & 1801 & 72.77\% & 0.39\%  & 26.83\% & 97.29\% & 100.00\% \\
        \methodBSO & P2           & 4 & 805  & 3220 & 678  & 2542 & 77.86\% & 0.37\%  & 21.77\% & 92.77\% & 99.92\%  \\
        \methodBSO & P3           & 4 & 284  & 1136 & 269  & 867  & 76.41\% & 0.09\%  & 23.50\% & 85.87\% & 100.00\% \\
        \methodBSO & P4           & 5 & 207  & 1035 & 195  & 840  & 81.55\% & 0.00\%  & 18.45\% & 41.03\% & 100.00\% \\
        \methodBSO & \accumulated &   & 2142 & 7929 & 1879 & 6050 & 76.50\% & 0.29\%  & 23.21\% & 88.19\% & 99.97\%  \\
        \methodGPT & P1           & 3 & 846  & 2538 & 737  & 1801 & 66.04\% & 0.51\%  & 33.45\% & 83.72\% & 99.83\%  \\
        \methodGPT & P2           & 4 & 805  & 3220 & 678  & 2542 & 75.71\% & 0.22\%  & 24.07\% & 83.33\% & 99.76\%  \\
        \methodGPT & P3           & 4 & 284  & 1136 & 269  & 867  & 75.62\% & 1.14\%  & 23.24\% & 81.04\% & 99.88\%  \\
        \methodGPT & P4           & 5 & 207  & 1035 & 195  & 840  & 83.09\% & 0.10\%  & 16.81\% & 34.36\% & 99.88\%  \\
        \methodGPT & \accumulated &   & 2142 & 7929 & 1879 & 6050 & 73.57\% & 0.43\%  & 26.01\% & 78.07\% & 99.82\%  \\
        \midrule
        \multicolumn{12}{c}{\textbf{First-order logic}}\\
        \midrule
        \methodB   & F1           & 9 & 2208 & 19872 & 1426 & 18446 & 25.65\% & 67.75\% & 6.60\% & 67.39\% & 74.73\% \\
        \methodB   & F2           & 8 & 1159 & 9272  & 916  & 8356  & 2.92\%  & 87.81\% & 9.26\% & 53.38\% & 91.18\% \\
        \methodB   & \accumulated &   & 3367 & 29144 & 2342 & 26802 & 18.42\% & 74.14\% & 7.45\% & 61.91\% & 79.86\% \\
        \methodBGP & F1           & 9 & 2208 & 19872 & 1426 & 18446 & 89.41\% & 3.26\%  & 7.33\% & 78.33\% & 75.11\% \\
        \methodBGP & F2           & 8 & 1159 & 9272  & 916  & 8356  & 76.41\% & 14.36\% & 9.23\% & 54.15\% & 98.10\% \\
        \methodBGP & \accumulated &   & 3367 & 29144 & 2342 & 26802 & 85.28\% & 6.79\%  & 7.93\% & 68.87\% & 82.27\% \\
        \methodBGO & F1           & 9 & 2208 & 19872 & 1426 & 18446 & 88.80\% & 3.26\%  & 7.94\% & 78.05\% & 74.89\% \\
        \methodBGO & F2           & 8 & 1159 & 9272  & 916  & 8356  & 80.18\% & 11.38\% & 8.44\% & 63.97\% & 97.42\% \\
        \methodBGO & \accumulated &   & 3367 & 29144 & 2342 & 26802 & 86.06\% & 5.84\%  & 8.10\% & 72.54\% & 81.92\% \\
        \methodBSP & F1           & 9 & 2208 & 19872 & 1426 & 18446 & 91.06\% & 0.37\%  & 8.56\% & 85.97\% & 75.95\% \\
        \methodBSP & F2           & 8 & 1159 & 9272  & 916  & 8356  & 89.00\% & 1.13\%  & 9.87\% & 62.45\% & 99.87\% \\
        \methodBSP & \accumulated &   & 3367 & 29144 & 2342 & 26802 & 90.41\% & 0.61\%  & 8.98\% & 76.77\% & 83.40\% \\
        \methodBSO & F1           & 9 & 2208 & 19872 & 1426 & 18446 & 91.48\% & 0.82\%  & 7.70\% & 88.50\% & 75.89\% \\
        \methodBSO & F2           & 8 & 1159 & 9272  & 916  & 8356  & 89.54\% & 0.77\%  & 9.70\% & 66.70\% & 99.81\% \\
        \methodBSO & \accumulated &   & 3367 & 29144 & 2342 & 26802 & 90.86\% & 0.80\%  & 8.34\% & 79.97\% & 83.35\% \\
        \methodGPT & F1           & 9 & 2208 & 19872 & 1426 & 18446 & 87.66\% & 3.95\%  & 8.39\% & 70.20\% & 72.02\% \\
        \methodGPT & F2           & 8 & 1159 & 9272  & 916  & 8356  & 85.69\% & 4.72\%  & 9.59\% & 61.03\% & 97.74\% \\
        \methodGPT & \accumulated &   & 3367 & 29144 & 2342 & 26802 & 87.03\% & 4.20\%  & 8.77\% & 66.61\% & 80.04\% \\
        \midrule
    \end{tabularx}
    \caption{%
        Accuracy of NLP-based methods in \emph{multi-class classification}. Each
        method is evaluated individually for each assignment, which is
        indicated in the id column. Id \accumulated is used for the aggregated
        values that can also be found in \cref{table:multi-class}.    
}\label{table:multi-class-evaluation}
\end{table}

\newcommand{\originalExerciseText}[2]{%
		\subsubsection{German version.}
		#2
}
\newcommand{\translatedExerciseText}[2]{%
		\subsubsection{English version.}
		#2
}
\newcommand{\translatedSolutionSpace}[1]{%
    \subsubsection{Solution space (English).}
    #1
}
\section{Assignments for Propositional Logic}\label{section:appendix-propositional-exercises}

We devised four assignments for propositional logic. For each, we provide German original used in our course as well as an English translation. We also provide solution space in English.

\subsection{Assignment: Lecture Participation}
Context: An exercise about a group of three friends visiting a lecture on logic.

\originalExerciseText{Lecture Participation}{%
    Bea, Kim und Wim hören dieses Semester Logik. Doch alle drei erscheinen nur unregelmäßig zur Vorlesung. In der Übung, in der alle drei zuverlässig anwesend sind, versuchen sie, Regelmäßigkeiten in ihrem Besuch der Vorlesung zu identifizieren. Sie befragen dazu ihre Kommilitonen, die sich nicht einig sind und die folgenden – zum Teil widersprüchlichen – Aussagen über den Vorlesungsbesuch von Bea, Kim und Wim äußern:
\begin{enumerate}
    \item Bea nimmt nur an der Logik-Vorlesung teil, wenn mindestens einer der anderen beiden daran teilnimmt.
    \item Mindestens zwei der drei Freunde nehmen an der Logik-Vorlesung teil.
    \item Wenn weder Bea noch Wim an der Vorlesung teilnehmen, dann nimmt Kim auch nicht teil.
    \item Keiner der drei nimmt an der Vorlesung teil.
\end{enumerate}%

Kannst du ihnen dabei helfen, diese Aussagen mit Hilfe aussagenlogischer Formeln zu modellieren? Identifiziere zunächst geeignete aussagenlogische Variablen und stelle anschließend aussagenlogischen Formeln auf.
}

\translatedExerciseText{Lecture Participation}{%
    Bea, Kim and Wim are studying logic this semester. However, all three of them only attend the lecture irregularly. In the tutorial sessions, in which all three are reliably present, they try to identify regularities in their attendance at the lecture. They ask their fellow students, who disagree and make the following -- sometimes contradictory -- statements about Bea, Kim and Wim's lecture attendance:
\begin{enumerate}
    \item Bea only attends the logic lecture if at least one of the other two attends.
    \item At least two of the three friends attend the logic lecture.
    \item If neither Bea nor Wim attend the lecture, then Kim doesn't attend either.
    \item None of the three attend the lecture.
\end{enumerate}%
Can you help them to model these statements using propositional formulas? First identify suitable propositional variables and then write propositional formulas.}

\translatedSolutionSpace{%
\begin{itemize}
    \item $\calV = \left\{
            \begin{tabular}{l l}
                $ (B,$&$ \text{Bea attends the logic lecture.}),$\\ 
                $ (K,$&$ \text{Kim attends the logic lecture.}),$\\ 
                $ (W,$&$ \text{Wim attends the logic lecture.})$ 
            \end{tabular}%
        \right\} $
    \item $ \calS = \{ \{ B, K, W \} \} $
\end{itemize}%
}

\subsection{Assignment: Movie Night}
Context: An exercise about a group of friends trying to figure out which films to watch.

\originalExerciseText{Movie Night}{%
Der kleine Tim möchte mit seinen Freunden einen Filmabend veranstalten. Essen und Getränke stehen schon bereit, nun müssen sie sich „nur“ noch darauf einigen, welche Filme sie schauen wollen. Zur Auswahl stehen die vier Filme: Der Pate, The Dark Knight, Star Wars: Episode IV und Forrest Gump. Nach einer längeren Diskussion haben die Freunde folgende Bedingungen an die Filmauswahl zusammengestellt:

\begin{enumerate}
    \item Die Gruppe schaut auf jeden Fall einen der vier Filme.
    \item Tim hat Forrest Gump schon recht häufig gesehen, deshalb wird er nur zustimmen, diesen Film zu schauen, wenn sie auch einen weiteren Film schauen.
    \item Aufgrund der ungeschriebenen Regel, dass schwarz gekleidete Personen mit übermenschlichen Fähigkeiten nur im Doppel auftreten dürfen, schauen sie The Dark Knight genau dann, wenn sie auch Star Wars: Episode IV schauen.
    \item Da der kleine Tim auch irgendwann schlafen muss, können sie nicht Star Wars: Episode IV schauen, wenn sie Der Pate und The Dark Knight schauen.
    \item Weiterhin einigen sie sich darauf, dass sie nur dann Der Pate nicht schauen, wenn sie nicht The Dark Knight oder nicht Star Wars: Episode IV schauen.
    \item Als letzte Bedingung stellen sie auf, dass sie Forrest Gump oder Star Wars: Episode IV schauen, wenn sie Der Pate sehen.
\end{enumerate}%
An dieser Stelle ruft der kleine Tim erfreut in die Runde, dass sich aus diesen Einschränkungen bereits ergibt, dass sie sowohl Der Pate als auch Forrest Gump aber keinen der anderen beiden Filme schauen.

Zeige, dass Tim recht hat, indem du die Situation aussagenlogisch modellierst und seine Aussage aus den Bedingungen folgerst.
}

\translatedExerciseText{Movie Night}{%
    Tim wants to have a movie night with his friends. Food and drinks are already prepared, now they “only” have to agree on the movies they want to watch. They can choose between the following four movies: The Godfather, The Dark Knight, Star Wars: Episode IV, and Forrest Gump. After a lengthy discussion, the friends have put together the following conditions for the movie selection:

    \begin{enumerate}
        \item The group will definitely watch one of the four films.
        \item Tim has seen Forrest Gump quite often, so he will only agree to watch this movie if they also watch another movie.
        \item Because of the unwritten rule that people dressed in black with superhuman abilities may only appear in pairs, they will watch The Dark Knight exactly when they also watch Star Wars: Episode IV.
        \item Since Tim eventually has to sleep, they will not be able to watch Star Wars: Episode IV if they watch The Godfather and The Dark Knight.
        \item They also agree that they will only not watch The Godfather if they do not watch The Dark Knight or not watch Star Wars: Episode IV.
        \item As a last condition they set up that they will watch Forrest Gump or Star Wars: Episode IV if they watch The Godfather.
    \end{enumerate}%
    At this point, Tim happily calls out to the group that these restrictions already imply that they watch both The Godfather and Forrest Gump but none of the other two movies.

    Show that Tim is right by modelling the situation with propositional formulas and inferring his statement from the conditions.
}

\translatedSolutionSpace{%
\begin{itemize}
    \item $\calV = \left\{ 
            \begin{tabular}{l l}
                $ (K,$&$ \text{The group watches The Dark Knight.}),$\\ 
                $ (P,$&$ \text{The group watches Der Pate.}),$\\ 
                $ (F,$&$ \text{The group watches Forrest Gump.}),$\\
                $ (S,$&$ \text{The group watches Star Wars: Episode IV.})$
            \end{tabular}%
    \right\}$
    \item $ \calS = \{ \{ K, P, F, S \} \} $
\end{itemize}%
}

\subsection{Assignment: Chat System}
Context: An exercise about a group of four friends trying to debug a faulty chat server.

\originalExerciseText{Chat System}{%
Sophie, Luke, Maja und Nino entwickeln ein neues Chat-System. Beim Testen fällt ihnen auf, dass einige Nachrichten bei der Übertragung verloren gehen. Um den entsprechenden Bug in ihrer Implementierung zu finden, schicken sie eine Reihe von Testnachrichten und beobachten ihr System dabei. Sie machen die folgenden Beobachtungen:

\begin{enumerate}
    \item Luke schaut in die Server-Logs und schlussfolgert: „Wenn Nino und Sophie jede Testnachricht erhalten haben, dann ich auch.“
    \item Maja vergleicht die Client-Logs miteinander und ist sich sicher: Sophie und Luke können nur dann beide alle Testnachrichten bekommen haben, wenn Maja selbst nicht alle erhalten hat.
    \item Sowohl Sophie als auch Nino stellen fest, dass sie jede Testnachricht erhalten haben.
\end{enumerate}%

Die vier vermuten nun, dass Maja nicht alle Testnachrichten erhalten hat. Zeige mit Mitteln der Aussagenlogik, dass diese Vermutung in der Tat korrekt ist und formal aus ihren Beobachtungen folgt.
Bevor du die Aussagen (1) bis (3) durch aussagenlogische Formeln modellieren kannst, musst du zunächst die dafür nötigen aussagenlogischen Variablen und deren intendierte Bedeutung festlegen.
}

\translatedExerciseText{Chat System}{%
Sophie, Luke, Maja and Nino are developing a new chat system. Whilst testing it, they notice that some messages get lost during transmission. In order to find the corresponding bug in their implementation, they send a series of test messages and observe their system. They make the following observations:
\begin{enumerate}
    \item Luke looks at the server logs and concludes: ``If Nino and Sophie have received every test message, then so have I.''
    \item Maja compares the client logs and is certain that Sophie and Luke can only have received all the test messages if Maja hasn't received all of them herself.
    \item Both Sophie and Nino realise that they have received every test message.
\end{enumerate}%

The four of them now assume that Maja has not received all the test messages. Use propositional logic to show that this assumption is in fact correct and follows formally from their observations.
Before you can model statements (1) to (3) using propositional formulas, you must first define the necessary propositional variables and their intended meaning.
}

\translatedSolutionSpace{%
\begin{itemize}
    \item $\calV = \left\{ 
            \begin{tabular}{l l}
                $ (S,$&$ \text{Sophie received all test messages.}),$\\ 
                $ (L,$&$ \text{Luke received all test messages.}),$\\ 
                $ (M,$&$ \text{Maja received all test messages.}),$\\
                $ (N,$&$ \text{Nino received all test messages.})$
            \end{tabular}%
    \right\}$
    \item $ \calS = \{ \{ S, L, M, N \} \} $
\end{itemize}%
}

\subsection{Assignment: Hardware System}
Context: An exercise about a faulty hardware system. In contrast to the other exercises, the five solutions here are each phrased differently; different expressions are used to state the faultiness of the five hardware components under investigation.

\originalExerciseText{Hardware System}{%
Die beiden Admins Paul und Fiona wollen einen defekten Laptop zum Laufen bringen. Sie starten ein Diagnoseprogramm, um einige Hardwarekomponenten auf ihre Funktionsfähigkeit zu testen. Nach mehreren Durchläufen können sie der Log-Datei des Diagnoseprogramms folgende Zusammenhänge entnehmen:

\begin{enumerate}
    \item Der Laptop stürzt genau dann ab, wenn ein Festplattenfehler auftritt.
    \item Die Netzwerkkarte und die Grafikkarte zeigen nicht beide ein fehlerhaftes Verhalten.
    \item Nur wenn die Grafikkarte ein fehlerhaftes Verhalten zeigt, ist weder der Laptop abgestürzt noch ein Festplattenfehler aufgetreten.
    \item Wenn der Laptop abstürzt und ein Festplattenfehler auftritt, dann klemmt das DVD-Laufwerk.
    \item Wenn die Grafikkarte ein fehlerhaftes Verhalten zeigt, dann klemmt das DVD-Laufwerk oder die Netzwerkkarte zeigt ein fehlerhaftes Verhalten.
\end{enumerate}%

Fiona vermutet, dass das DVD-Laufwerk auf jeden Fall klemmt. Überzeuge Paul davon, dass Fionas Vermutung stimmt, indem du zunächst die Situation aussagenlogisch modellierst und dann zeigst, dass Fionas Vermutung aus den gegebenen Zusammenhängen folgt. Bevor du die Aussagen (1) bis (5) durch aussagenlogische Formeln modellieren kannst, musst du zunächst die dafür nötigen aussagenlogischen Variablen und deren intendierte Bedeutung festlegen.
}

\translatedExerciseText{Hardware System}{%
The two admins Paul and Fiona want to get a faulty laptop up and running. They start a diagnostic programme to test the functionality of some hardware components. After several runs, they can see the following correlations in the log file of the diagnostic programme:

\begin{enumerate}
    \item The laptop crashes exactly when a hard drive error occurs.
    \item The network card and the graphics card do not both show faulty behaviour.
    \item Only if the graphics card shows faulty behaviour, neither the laptop has crashed nor a hard disk error has occurred.
    \item If the laptop crashes and a hard drive error occurs, then the DVD drive is stuck.
    \item If the graphics card shows faulty behaviour, then the DVD drive is stuck or the network card is showing faulty behaviour.
\end{enumerate}%

Fiona conjectures that the DVD drive is stuck. Convince Paul that Fiona's assumption is correct by first modelling the situation using propositional logic and then showing that Fiona's assumption follows from the given contexts. Before you can model statements (1) to (5) using propositional formulas, you must first define the necessary propositional variables and their intended meaning.
}

\translatedSolutionSpace{%
\begin{itemize}
    \item $\calV = \left\{ 
            \begin{tabular}{l l}
                $ (L,$&$ \text{The laptop crashes.}),$\\ 
                $ (F,$&$ \text{A hard drive error occurs.}),$\\ 
                $ (N,$&$ \text{The network card shows faulty behaviour.}),$\\
                $ (G,$&$ \text{The graphics card shows faulty behaviour.}),$\\
                $ (D,$&$ \text{The DVD drive is stuck.})$
            \end{tabular}%
    \right\}$
    \item $ \calS = \{ \{ L, F, N, G, D \} \} $
\end{itemize}
}

\section{Assignments for First-order Logic}\label{section:appendix-fo-exercises}

We devised two assignments for first-order logic. For each, we provide German original used in our course as well as an English translation. We also provide solution space in English.

Assignments in the \Iltis educational support system are specified in XML. We provide these specifications of these assignments in XML.

As sketched in the main part of the paper, instructors can provide custom feedback texts targeting specific cases. Those cases are split into \emph{faults}, which are used to provide targeted feedback for selected mistakes, and \emph{suggestions} informing students of how they could have improved their vocabulary. While the display of faults prevents the task from being complete, students may choose to ignore suggestions and continue with their own vocabulary.

In our assignments, after designing a first-order vocabulary, students are asked to write first-order formulas over their chosen vocabulary $ V $. An instructor has to provide the solution formulas in advance, say over vocabulary $ V^* $.
To facilitate the translation of student formulas over $ V $ into formulas over $ V^* $ -- e.g. to check if the student formula is equivalent to the solution formula -- instructors need to provide ``translations'' for symbols not in $ V^* $. For example, if a student chose $ P $ (unary relation symbol) rather than $ p $ (constant symbol) in their vocabulary, each occurrence of $ P(x) $ in their formulas will be replaced by $ x = p $, see first XML below. In that assignment, there is also a suggestion to use a constant symbol instead. Since permutations of arguments for symbols of arity at least two occurs regularly, our system translates them automatically, provided the permutation is indicated in the corresponding description (see e.g.\ symbol $ F $).

\newcommand{\paramU}{\ensuremath{u}\xspace}
\newcommand{\paramV}{\ensuremath{v}\xspace}

\subsection{Assignment: Book Collection}~

\originalExerciseText{Book Collection}{
    Im Folgenden sollen Eigenschaften einer Sammlung von Büchern und ihren Autoren modelliert werden. Jedes Buch hat genau einen Autor. Manche der Autoren sind Mathematiker. Einige der Bücher widerlegen andere Bücher. Manche Bücher behandeln Logik. Das Buch Principia Mathematica darf in der Sammlung nicht fehlen.

    Hinweis: In der Realität haben viele Bücher mehrere Autoren. Wir nehmen hier aber vereinfachend an, dass jedes Buch genau einen Autor hat.

    Außerdem gelten folgende Eigenschaften:

    \begin{enumerate}
        \item Die Principia Mathematica wurde von einem Mathematiker geschrieben und behandelt Logik.
        \item Bücher über Logik werden nur von Büchern widerlegt, die auch Logik behandeln.
        \item Es gibt ein Buch, das weder von einem Mathematiker verfasst wurde, noch von einem anderen Buch widerlegt wird.
        \item Jeder Autor hat ein Buch geschrieben, das Logik behandelt oder von einem anderen Buch widerlegt wird. 
    \end{enumerate}%

    Gib anhand des obigen Textes natürlichsprachliche Beschreibungen der Symbole einer Struktur an, mit der man dieses Szenario möglichst natürlich modellieren kann. Das Universum der Struktur enthält dabei alle Bücher und Autoren.
}

\translatedExerciseText{Book Collection}{
    In the following, the properties of a collection of books and their authors will be modelled. Each book has exactly one author. Some of the authors are mathematicians. Some of the books refute other books. Some books deal with logic. The book Principia Mathematica must not be missing from the collection.

    Note: In reality, many books have several authors. However, for the sake of simplicity, we assume that each book has exactly one author.

    The following properties also apply:
    \begin{enumerate}
        \item The Principia Mathematica was written by a mathematician and deals with logic.
        \item Books on logic are only refuted by books that also deal with logic.
        \item There is a book that was neither written by a mathematician nor refuted by another book.
        \item Every author has written a book that deals with logic or is refuted by another book.
    \end{enumerate}%

    Using the above text, give natural language descriptions of the symbols of a structure that can be used to model this scenario as naturally as possible. The domain of the structure contains all books and authors.
}

\translatedSolutionSpace{%
    \begin{itemize}
        \item $\calV = \left\{ 
                \begin{tabular}{l p{60mm}}
                    $ (B(\paramU),$          & $ \text{\paramU is a book.}),$\\
                    $ (A(\paramU),$          & $ \text{\paramU is an author.}),$\\
                    $ (M(\paramU),$          & $ \text{Author \paramU is a mathematician.}),$\\
                    $ (L(\paramU),$          & $ \text{Book \paramU deals with logic.}),$\\
                    $ (R(\paramU, \paramV),$ & $ \text{Book \paramU refutes book \paramV.}),$\\
                    $ (f(\paramU),$          & $ \text{Author of book \paramU}),$\\
                    $ (F(\paramU, \paramV),$ & $ \{\text{\paramU is the author of book \paramV.}\newline \text{~\,Book \paramU was written by author \paramV.}\}),$\\
                    $ (p,$                   & $ \text{The book Principia Mathematica}),$\\
                    $ (P(\paramU),$          & $ \text{\paramU is the book Principia Mathematica.})$
                \end{tabular}%
        \right\}$
        \item $ \calS = \left\{ 
                \begin{tabular}{l}
                    $ \{ B, A, M, L, R, f, p\},$\\
                    $ \{ B, A, M, L, R, f, P \},$\\ 
                    $\{ B, A, M, L, R, F, p \},$\\
                    $\{ B, A, M, L, R, F, P \} $
                \end{tabular}%
        \right\} $
        \item $ \varphi_\calS = B \land A \land M \land L \land R \land (f \lor F) \land (p \lor P) $ with redundancies $ \{ f, F \} $ and $ \{ p, P \} $
    \end{itemize}%
}

\subsubsection{Specification of the Solution Space in XML (English):}~

\begin{lstlisting}
<Symbols>
  <Relation symbol="/*!\textcolor{orange}{$B$}!*/" arity="1">
    <Description>/*!$\paramU$!*/ is a book.</Description>
  </Relation>

  <Relation symbol="/*!\textcolor{orange}{$A$}!*/" arity="1">
    <Description>/*!\paramU!*/ is an author.</Description>
  </Relation>

  <Relation symbol="/*!\textcolor{orange}{$M$}!*/" arity="1">
    <Description>Author /*!\paramU!*/ is a mathematician.</Description>
  </Relation>

  <Relation symbol="/*!\textcolor{orange}{$L$}!*/" arity="1">
    <Description>Book /*!$\paramU$!*/ deals with logic.</Description>
  </Relation>

  <Relation symbol="/*!\textcolor{orange}{$R$}!*/" arity="2">
    <Description>Book /*!$\paramU$!*/ refutes book /*!$\paramV$!*/.</Description>
  </Relation>

  <Function symbol="/*!\textcolor{orange}{$f$}!*/" arity="1">
    <Description>Author of book /*!$\paramU$!*/</Description>
  </Function>

  <Relation symbol="/*!\textcolor{orange}{$F$}!*/" arity="2">
    <Description>/*!$\paramU$!*/ is the author of book /*!$\paramV$!*/.</Description>
    <Description permutation="/*!\textcolor{orange}{$\paramU,\paramV$}!*/">
      Book /*!$\paramU$!*/ was written by author /*!$\paramV$!*/.
    </Description>
    <Translation>/*!$u = f(v) \land B(v)$!*/</Translation>
  </Relation>

  <Constant symbol="/*!\textcolor{orange}{$p$}!*/" arity="0">
    <Description>The book Principia Mathematica</Description>
  </Constant>

  <Relation symbol="/*!\textcolor{orange}{$P$}!*/" arity="1">
    <Description>
      /*!$\paramU$!*/ is the book Principia Mathematica.
    </Description>
    <Translation>/*!$\paramU = p$!*/</Translation>
  </Relation>
</Symbols>

<Faults>
  <Fault when="/*!\textcolor{orange}{$\lnot(B \land A)$}!*/">
    Think again about what types of elements there are in
    this scenario. For a complete characterisation, your
    signature should contain a unary relation for each of
    these types.
  </Fault>

  <Fault when="/*!\textcolor{orange}{$\lnot(M \land (F \lor f) \land (P \lor p) \land L)$}!*/">
    Look at the following statement. Can you already model
    it with your signature?
    <blockquote>
      The Principia Mathematica was written by a
      mathematician and deals with logic.
    </blockquote>
  </Fault>

  <Fault when="/*!\textcolor{orange}{$\lnot(R \land L)$}!*/">
    Look at the following statement. Can you already model
    it with your signature?
    <blockquote>
      Books on logic are only refuted by books that also
      deal with logic.
    </blockquote>
  </Fault>
</Faults>

<Suggestions>
  <Suggestion when="/*!\textcolor{orange}{$F \land \lnot f$}!*/">
    Instead of a relation, you could use a function here,
    since each element of the structure has (at most) one
    author.  
  </Suggestion>

  <Suggestion when="/*!\textcolor{orange}{$P \land \lnot p$}!*/">
    Instead of a relation, you could use a constant here, 
    as Principia Mathematica occurs exactly once in the
    structure.
  </Suggestion>
</Suggestions>

<CompletenessCondition>
  /*!$B \land A \land M \land L \land W \land (F \lor f) \land (P \lor p)$!*/
</CompletenessCondition>

<Redundancies>
  <Set>/*!$F,f$!*/</Set>
  <Set>/*!$P,p$!*/</Set>
</Redundancies>
\end{lstlisting}

\subsection{Assignment: Faculty Conference}~

\originalExerciseText{Faculty Conference}{%
    Im Zuge der Digitalisierung der Hochschulverwaltung soll die Konferenzteilnahme der Mitarbeitenden der Fakultät Informatik in einer Datenbank erfasst werden. Folgende Rahmenbedingungen sind gegeben:

    Die Fakultät besteht aus mehreren Arbeitsgruppen. Diese bestehen jeweils aus mehreren Personen, die an dieser Arbeitsgruppe arbeiten. Diese bezeichnen wir als Mitglieder der Arbeitsgruppe. Jede Person ist Mitglied genau einer Arbeitsgruppe. Eine von vielen Arbeitsgruppen ist die Arbeitsgruppe "Logik und formale Verifikation". Mitglieder von Arbeitsgruppen können Konferenzen besuchen und dort eine wissenschaftliche Arbeit vorstellen, und auch an Konferenzen teilnehmen, ohne dort eine wissenschaftliche Arbeit vorzustellen.

    Gib anhand des obigen Textes natürlichsprachliche Beschreibungen der Symbole einer Struktur an, mit der man dieses Szenario möglichst natürlich modellieren kann. 
    Das Universum der Struktur enthält Arbeitsgruppen, Personen, und Konferenzen.
}

\translatedExerciseText{Faculty Conference}{%
    As part of the digitalisation of the university administration, the conference attendance of employees of the Faculty of Computer Science is to be recorded in a database. The following framework conditions are in place:

    The faculty consists of several working groups. Each of these consists of several people who work on this working group. We refer to these as members of the working group. Each person is a member of exactly one working group. One of many working groups is the working group ``Logic and Formal Verification''. Members of working groups can attend conferences and present a scientific paper there, and can also participate in conferences without presenting a scientific paper there.

    Using the above text, give natural language descriptions of the symbols of a structure that can be used to model this scenario as naturally as possible.
    The domain of the structure contains working groups, persons and conferences.
}

\translatedSolutionSpace{%
    \begin{itemize}
        \item $\calV = \left\{ 
                \begin{tabular}{l l}
                    $ (W(\paramU),$          & $ \text{\paramU is a working group.}),$\\
                    $ (P(\paramU),$          & $ \text{\paramU is a person.}),$\\
                    $ (C(\paramU),$          & $ \text{\paramU is a conference.}),$\\
                    $ (A(\paramU, \paramV),$ & $ \text{Person \paramU attends the conference \paramV without presenting.}),$\\
                    $ (N(\paramU, \paramV),$ & $ \text{Person \paramU attends the conference \paramV.}),$\\
                    $ (T(\paramU, \paramV),$ & $ \text{Person \paramU presents at conference \paramV.}),$\\
                    $ (g(\paramU),$          & $ \text{The working group of person \paramU}),$\\
                    $ (l,$                   & $ \text{The working group ``Logic and Formal Verification''})$
                \end{tabular}%
        \right\}$
        \item $ \calS = \{ \{ W, P, C, A, T, g, l \}, \{ W, P, C, N, T, g, l \} \} $
        \item $ \varphi_\calS \df W \land P \land C \land (A \lor N) \land T \land g \land l $ with redundancies $ \{ A, N \} $
    \end{itemize}%
}

 \end{document}